\documentstyle[12pt,aasms4]{article}
\input{psfig}

\begin{document}

\title{Flows on Scales of 150 Mpc?}
\author{K. R. M\"uller}
\affil{Dept of Physics and Astronomy, Dartmouth College, Hanover
NH 03755-3528 and European Southern Observatory,
Karl Schwarzschild Str. 2, D-85748 Garching bei M\"unchen, Germany}
\authoremail{kmueller@eso.org}

\author{W. Freudling}
\affil{European Southern Observatory and Space Telescope -- European
Coordinating Facility, Karl Schwarzschild Str. 2, D-85748 Garching bei
M\"unchen, Germany}
\and                         
\author{R. Watkins \&  G. Wegner}
\affil{Dept of Physics and Astronomy, Dartmouth College,
Hanover NH 03755-3528}

\begin{abstract}

In order to investigate the reality of large-scale streaming motion 
on scales of up to 150~Mpc, we have studied the peculiar motions 
of $\sim$ 200 early-type galaxies in three directions of the South
Equatorial Strip, at distances out to $\sim$ 20~000~km~s$^{-1}$.
The new APM South Equatorial Strip Catalog
($-17^{\circ}.5 < \delta < +2^{\circ}.5$)
was used to select the sample of field galaxies in three
directions: (1) 15h10 -- 16h10; (2) 20h30 -- 21h50; (3) 00h10 -- 01h30.
New R-band CCD photometry and spectroscopic data for the galaxies are used.
The Fundamental Plane distance-indicator relation is calibrated
with Coma cluster data, and a correction for inhomogeneous
Malmquist bias is applied to the distance estimates.
A linear bulk-flow model is fitted to the peculiar velocities in
the sample regions and the results do not reflect the bulk flow
observed by Lauer and Postman (LP).  
Accounting for the difference in geometry between
the galaxy distribution in the three regions and the LP clusters
confirms the disagreement; assuming a low-density CDM power
spectrum, we find that the observed bulk flow of the galaxies in 
our sample excludes the LP bulk flow at the $99.8\%$ confidence
level.

\end{abstract}     


\section{Introduction}

There is strong observational evidence for the existence of
large-scale flows in the local universe, induced by gravity (see
\markcite{sw95}Strauss \& Willick 1995).  The dipole anisotropy of the
cosmic microwave background (CMB) radiation provides a natural
velocity reference frame for the analysis of galaxy motions.  The
dipole anisotropy, determined from COBE, implies that the Local Group
(LG) moves with respect to the CMB rest frame at 627 $\pm$ 22
km~s$^{-1}$ towards $ l = 276 \pm 3^{\circ}$, $b = +30 \pm 3^{\circ}$
(\markcite{kog93}Kogut $et$ $al.$ 1993).  If this has a kinematic
origin then, sufficiently far away, the peculiar velocities should
converge to the CMB frame. Indeed, the observed LG motion relative to
the restframe defined by galaxies within 6000 km/sec points towards
the Hydra Centaurus-Great Attractor region points, which is within
$\sim 40^{\circ}$ of the CMB dipole direction.

The observed LG motion towards the Hydra Centaurus-Great
Attractor region points within $\sim 40^{\circ}$
of the CMB dipole direction. However, until recently only the
region within about 6000 km~s$^{-1}$ had been well sampled.

Until now, the only studies which have reported measurements of 
the velocity field as far out as 15~000 km~s$^{-1}$ are those of
\markcite{lp94}Lauer \& Postman (LP) (1994),
using brightest cluster galaxies as distance
estimators, and \markcite{ri95}Riess $et$ $al.$ (1995),
using Type Ia supernovae.
LP checked the convergence of the LG dipole motion
to the CMB dipole, with a surprising result: a strong signature
of a very large-scale bulk flow was seen, with an amplitude of
689 $\pm$ 178 km~s$^{-1}$ in the direction $l = 343^{\circ},
b = +52^{\circ}$.                                      
The LP study implies that the local rest frame fails to converge
to the CMB frame, even in regions with radii 
$\sim$ 15~000 km~s$^{-1}$.
A bulk flow with the statistical significance of this result
rules out a whole series of cosmological models at the $> 95$\,\%
confidence level (\markcite{fw94}Feldman \& Watkins
1994; \markcite{st95}Strauss $et$ $al.$ 1995); the LP result 
is in disagreement with all viable models at present.

The LP sample extended to 15~000 km~s$^{-1}$, with an effective
depth of $\sim 8000$ km~s$^{-1}$.
Therefore, the logical next step was to compare the LP result with
peculiar velocities as found from applying the Tully-Fisher and
Fundamental Plane methods to galaxies extending further out than
any previous peculiar velocity studies.
Neither field nor cluster spiral galaxies within 8000~km~s$^{-1}$ both
show evidence of such a motion 
\markcite{gi96}\markcite{gi98a}\markcite{g98b}Giovanelli $et$ $al.$ 
(1996, 1998a, 1998b).

In this work, we analyze new and independent measurements of the
peculiar velocity field of elliptical field galaxies at a depth
similar to that of LP. Most of the galaxies are within 10~000 
km~s$^{-1}$, with some as far out as 20~000 km~s$^{-1}$.
A sample of 179 early-type galaxies in three selected 
regions was used to investigate peculiar motions.
The first sample region is about 20$^{\circ}$ from the direction 
of the LP bulk flow.
The second region is almost perpendicular to the first direction,
and the third is in a direction on the opposite side of the sky
from the first, close to the direction of the Perseus-Pisces region
and the South Galactic Pole.
                                             
\section{Sample Selection and Observations}

Galaxies were selected from the new APM South Equatorial Strip 
Catalog, made available by Somak Raychaudhury prior to publication
(\markcite{ra98}Raychaudhury $et$ $al.$ 1998).
The South Equatorial Strip, which lies between declination 
$-17^{\circ}.5$ and $+2^{\circ}.5$,
is an uncharted region in the velocity field, because previously
no good galaxy catalog existed for this region,
and consequently the peculiar motions of galaxies in this strip 
had never been mapped.

The present sample has well-defined uniform selection criteria.
Starting with the APM South Equatorial Strip Catalog with a magnitude
limited of $b_j = 17.0$ mag, candidate galaxies in the three regions
were examined on the POSS plates and later on CCD images to verify the
morpholotical type. This resulted in a sample of E/S0 galaxies with a
magnitude cut which is virtually complete to Kron-Cousins
$R=14.0$. The completness drops for fainter magnitudes, galaxies down
to $R$=15.05 are included.

Spectra and R-band CCD images of galaxies in the three sample regions
were collected during a series of ten observing runs between June 1993
and September 1995. Observations were made at the 1.3~m McGraw-Hill telescope
and the 2.4~m Hiltner telescope of the Michigan-Dartmouth-M.I.T. Observatory,
and also at the 4.4~m Multiple Mirror Telescope on Mount Hopkins, Arizona.

The galaxy images were processed and photometric parameters were
derived with the procedures and programs described by
\markcite{freud91}Freudling et al. (1991)
and Saglia $et$ $al.$ (1997a,b).  Azimuthally averaged surface
brightness profiles were fitted as the sum of bulge and disk
components, and a correction for seeing was applied (see
\markcite{sag97}Saglia $et$ $al.$ 1997a).  The total magnitude
$m_{tot}$ and the half-light parameters $r_{1/2}$ (half-light radius),
$\mu_{1/2}$ (mean surface brightness at $r_{1/2}$), and
$<\!\!\mu\!\!>_{1/2}$ (mean surface brightness within $r_{1/2}$) were
derived.

Spectra were extracted from the spectroscopic observations using the
procedures outlined in Wegner $et$ $al.$ (1998). The median S/N per
$\AA$ is 23. The instrumental resolution (FWHM) is $\sim4 \AA$, and
the spectrograph resolution (dispersion) is $\sim$100 km/s.  Fourier
cross-correlation analysis was used to determine values of redshift
and velocity dispersion $\sigma$ from logarithmically-rebinned
spectra.  Velocity dispersions were corrected for aperture effects.
The final data as well as full details of the observations, data
reduction, and analysis are given in \markcite{mu97}M\"uller (1997)
and
\markcite{mu98}M\"uller $et$ $al.$ (1998).

\section{Peculiar Velocities}

The Fundamental Plane (FP) distance-indicator relation was calibrated
with data from 40 galaxies in the Coma cluster. Care was taken to use
identical instruments and data reduction procedures. Objects were
chosen to be E/S0 using earlier studies ($e.g$ J$\o$rgensen $et$ $al.$
1993; Lucey $et$ $al.$ 1991) and lie mostly within about $0.5^\circ$
of the point midway between NGC 4889 and NGC 4874.  The coefficients
were fitted by minimizing orthogonal residuals from the plane (as in
\markcite{jo96}J$\o$rgensen $et$ $al.$ 1996).  Monte Carlo simulations
were used to determine the bias in the coefficients, and a correction
for incompleteness was made following a similar procedure
to \markcite{gi97}Giovanelli $et$ $al.$ (1997).

The FP for Coma, determined from these data, is best described by the
relation
\begin{displaymath}
\log r_{1/2} = 1.247  \log \sigma + 0.348 <\!\!\mu\!\!>_{1/2} - \, 8.815,
\end{displaymath}
with a measured scatter equivalent to a distance uncertainty of 
19.4\,\%.  Coma is used as the reference cluster for the calibration 
of the distance-indicator relation.  Coma has a radial velocity of 
$\simeq$ 7200 km~s$^{-1}$ in the CMB reference frame.  The zero-point 
for the FP relation should be determined from a sample with no radial 
peculiar motion; in the case of Coma, the peculiar velocity is 
consistent with zero in the CMB frame, as shown by several authors 
(\markcite{fa89}Faber $et$ $al.$ 1989; \markcite{} Lucey $et$ $al.$ 
1991; \markcite{jo96}J$\o$rgensen $et$ $al.$ 1996; 
\markcite{gi97}Giovanelli $et$ $al.$ 1997; \markcite{sco97}Scodeggio 
$et$ $al.$ 1997).  The uncertainty in the adopted value for the 
peculiar velocity of Coma can conservatively be estimated to about 
$\pm$ 250 km~s$^{-1}$, but the actual choice of the velocity zeropoint 
has little effect on the results of the current analysis as explained 
below.

A cosmological correction was made to transform the derived distances
from diameter distances into true distances.  The sample was corrected
for inhomogeneous Malmquist bias, which corrects for the effect of
fluctuations in the galaxy density along the line of sight ({\it e.g.}
\markcite{w91}Willick 1991;
\markcite{f95}Freudling $et$ $al.$ 1995). These density fluctuations
were estimated from the new IRAS ``PSC-z'' 0.6 Jy redshift survey
(\markcite{saun94}\markcite{saun98}Saunders $et$ $al.$ 1994, 1998). 
This correction changes the  bulk flow component 
constrained by our data (see below) by approximately 30 km$~s^{-1}$.

The final sample contained only the galaxies with photometric and
spectroscopic data both of good quality: 50 galaxies in Region~1, 77
in Region~2, and 52 in Region~3.  The mean radial peculiar velocities
have been calculated in bins of inferred distance of 2000~km~s$^{-1}$.
The results are shown in Figures 1.  In these plots all galaxies in a
bin were replaced with one point at the distance of the center of the
bin, with the average value of the peculiar velocity in that bin.

\placefigure{bins}

\section{Analysis}

As a first approach to analyzing the peculiar motions, a linear
bulk-flow model was fitted to the data in each sample region, in
order to measure the radial component of the bulk
motion. Each data point was assigned a standard deviation according
to its distance error, and a weighted average peculiar velocity
$\overline{u}$ was determined from the galaxies in each region.
The estimated error $\sigma_{\overline{u}}$ of the weighted mean was
also found. All available data points were included in the fit; no
binning was used.

The results of the weighted fits were ($\overline{u} \pm
\sigma_{\overline{u}}$): ($+225 \pm 199$) km~s$^{-1}$ in Region~1;
($+145 \pm 162$) km~s$^{-1}$ in Region~2; and ($+468 \pm 197$)
km~s$^{-1}$ in Region~3.  The numbers of galaxies with inferred
distances $v_{corr} > 10~000$~km~s$^{-1}$ are small, and also the
distance errors are larger at those distances. Therefore the most
distant galaxies do not have a big effect on the weighted averages.
The few nearby galaxies, with small distance errors, have a large
effect on the weighted fits. If only the galaxies between 5000 and
15~000~km~s$^{-1}$ are used, the fits give the following results:
($+146 \pm 243$) km~s$^{-1}$ in Region~1; ($-9 \pm 228$) km~s$^{-1}$
in Region~2; and ($+340 \pm 307$) km~s$^{-1}$ in Region~3.  This means
that the resulting mean velocities are not statistically significant,
and the radial peculiar velocities are indistinguishable from zero.
At first glance, then, the results of this study appear to be
inconsistent with the hypothesis of a large bulk motion as suggested
by the LP results.

However, a simple direct comparison between the results for the
mean velocities in the sample regions and the flows expected
according to the result of LP can be misleading.  This is
because both the mean velocities and the LP bulk flow contain
contributions from the incomplete cancelation of velocity modes 
whose wavelengths are much smaller than the survey scale 
(\markcite{wf95}Watkins \& Feldman 1995).  Since the surveys 
have different geometries, these contributions will
be different for the two surveys.  It is therefore possible that
contributions from smaller-scale velocity modes could be the cause
of the observed disagreement; two studies of the same 
Universe and the same velocity field could produce different 
results for this reason.

We tested this possibility by calculating the covariance matrix for
the {\it difference} of the bulk flow of the sample taken as a       
whole (the galaxies in all three sample regions), $\vec U_E$, 
and the bulk flow observed by LP, $\vec U_{LP}$,
\begin{displaymath}
R_{ij}= \langle\  (\vec U_E - \vec U_{LP})_i\
 (\vec U_E - \vec U_{LP})_j\ \rangle .
\end{displaymath}
The covariance matrix takes into account contributions to the
measured bulk flow from both actual galaxy velocities and
measurement errors; it also properly accounts for correlations
between different components of the bulk flow.  We calculated
$R_{ij}$ using the method outlined in \markcite{wf95}Watkins \& 
Feldman (1995).
We use the maximum likelihood estimator for the bulk flow as given
in \markcite{k88}Kaiser (1988):
\begin{displaymath}
(\vec U)_i = A_{ij}^{-1}\sum_q\
{\hat r_{q,j}\ v_q\over (\sigma_q^2 + \sigma_*^2)},
\end{displaymath}
where
\begin{displaymath}
A_{i,j}= \sum_q\ {\hat r_{q,i}\ \hat r_{q,j}\over (\sigma_q^2 + \sigma_*^2)}.
\end{displaymath}
Here galaxies are labeled by an index $q$ and have positions $r_q$ and 
line-of-sight peculiar velocities $v_q$ with measurement uncertainties 
$\sigma_q$.  We assume $\sigma_q$ to be $19\%$ of a galaxy's distance 
for individual galaxies in our sample; for the LP clusters we 
calculate individual distance errors based on the quoted values of 
$\alpha_{C}$ (see \markcite{lp94}Lauer \& Postman 1994).  An additional 
uncertainty $\sigma_* = 350$ km~s$^{-1}$ is included to account for 
deviations from the linear flow field owing to nonlinear effects.  In 
practice, the specific value used for $\sigma_*$ changes the results 
of our analysis very little; most of the objects we are considering 
are at large enough distance that the errors in their velocities are 
much larger than this additional term.

To calculate the covariance matrix $R_{ij}$, one must assume a model
for the power spectrum, $P(k)$. Most popular models for $P(k)$ are
inconsistent with the LP result (\markcite{fw94}Feldman \& Watkins
1994;
\markcite{st95}Strauss $et$ $al.$ 1995).
However, since a large-scale flow should contribute equally to the bulk 
flows measured by each survey, the difference $\vec U_E- \vec U_{LP}$ should 
be almost entirely due to errors in the velocity measurements and to 
velocity modes on scales that are smaller than that of the surveys.  
On these scales, several  types of measurements have shown 
consistency with a low-density Cold Dark Matter ($\Omega_m=0.3$,$\Gamma=0.21$) 
model (LDCDM) normalized to COBE (see, {\it e.g.}, \markcite{sw95}Strauss
\& Willick 1995).  We used this model for our
representative $P(k)$; we also performed the analysis    
for a standard ($\Omega=1$,$\Gamma=0.5$) CDM model (SCDM) for
comparison, even though this model has been shown to be in disagreement
with several observations.

Using the model for $P(k)$, we calculated the covariance
matrix and the $\chi^2$ for the measured
difference between the bulk flows,
\begin{displaymath}
\chi^2 = \sum_{ij}(\vec U_E - \vec U_{LP})_i\ R^{-1}_{ij}
(\vec U_E - \vec U_{LP})_j.
\end{displaymath}
This $\chi^2$ can in turn be used to determine
the probability that the observed difference in bulk flow could
have arisen in a Universe with the assumed power spectrum.

Since the three survey regions lie roughly in the plane of the 
celestial equator, we are in practice restricted to measuring only two 
components of the bulk flow.  By diagonalizing the error matrix for 
the survey we determined that the two components of the bulk flow that 
can be measured with reasonable accuracy lie in a plane perpendicular 
to the direction $l= 114^\circ$, $b=23^\circ$ (about $10^\circ$ from 
the north pole of the celestial sphere).  In this plane, we find the 
bulk flow for the galaxies in the three regions to lie in the 
direction $l=49^\circ$, $b=-45^\circ$ with magnitude $312$ km~s$^{-1}$ 
(measured in the CMB frame).  Each of the two components of this flow 
have an uncertainty of approximately $180$km~s$^{-1}$, so our 
result is consistent with the frame defined by our 
galaxy sample being at rest.

This flow is quite consistent with expectations calculated assuming 
the LDCDM model.  The component of the LP bulk flow in this 
plane lies in the direction $l= 349^\circ$, $b= 54^\circ$ 
with magnitude $822$ km~s$^{-1}$.  (Note that the magnitude of 
the LP bulk flow given here is somewhat larger than the 
$689$ km~s$^{-1}$ reported by LP; their smaller value has been 
corrected for ``error bias'', whereas we are interested in the 
uncorrected maximum likelihood value of the bulk flow.) The two 
observed bulk flow vectors are separated by an angle of
$110^\circ$.  The component of the difference vector $\vec U_E-
\vec U_{LP}$ in the plane of interest has a magnitude of  
$982$ km~s$^{-1}$.

For the LDCDM model, the two measured components of $\vec
U_E-\vec U_{LP}$ give $\chi^2=12.7$ for two degrees of
freedom.  Therefore the possibility of velocity errors and 
small-scale velocity modes in the LDCDM model producing the 
observed difference $\vec U_E-\vec U_{LP}$ 
can be ruled out at the $99.8\%$ confidence level.  For
comparison, the SCDM model gives $\chi^2=8.5$, which corresponds 
to a $98.6\%$ confidence level.  The lower $\chi^2$ for the SCDM
model results from two factors.  First, the SCDM model has
relatively more of its power on smaller scales, where the modes are 
more likely to contribute to the disagreement between the two
observed bulk flows.  Second, since the velocity power
spectrum is proportional to $\Omega_m^{0.6}\sigma_8$, a larger
$\Omega_m$ will generally lead to all velocity modes having higher
amplitude.  However, even in the SCDM model, it is unlikely that the
observed difference $\vec U_E-\vec U_{LP}$ could arise from 
small-scale velocity modes.                                

The question arises as to how a different choice of the velocity 
zeropoint of our fundamental plane, or equivalently, an assumption of 
a nonzero peculiar velocity for the Coma cluster, would effect these 
results.  Changing the velocity zeropoint has the effect of adding or 
subtracting a percentage of each galaxy's redshift to its peculiar 
velocity.  This in turn changes the bulk flow of the sample by a 
vector whose magnitude is proportional to the assumed value of Coma's 
peculiar velocity, $V_{Coma}$.  The constant of proportionality and 
the direction of this vector depends on the geometry of the sample and 
its assumed velocity errors; for example, we would expect the change 
in bulk flow to be small for a sample whose galaxies are distributed 
isotropically.  For our sample, the addition to the bulk flow in the 
plane of interest has a magnitude of $1.18\ V_{Coma}$ and points in 
the direction $l=34^{\circ}$, $b=-22^{\circ}$, nearly 
perpendicular to the LP bulk flow.  This can be understood by noting 
that the contributions to this vector from Regions 1 and 3, which lie 
roughly parallel to the LP bulk flow and in opposite directions, 
approximately cancel, so that the dominant contribution is from 
Region~2, which is roughly perpendicular to the LP bulk flow.  The 
fact that the change in the bulk flow of the sample has only a small 
component in the direction of the LP bulk flow implies that changing 
the velocity zeropoint is unlikely to significantly improve their 
agreement.

To study the effect of changing the velocity zeropoint on our results, 
we have calculated the confidence level at which our sample rules out 
the LP bulk flow as a function of $V_{Coma}$.  We found that the 
confidence level was minimum for $V_{Coma}= -430$ km~s$^{-1}$, 
corresponding to $99.6\% $ and $97.9\% $ for the LDCDM and SCDM 
models respectively.  It should be noted that the assumption of a 
positive value for $V_{Coma}$ ,as found by LP, results in the LP bulk 
flow being ruled out at higher confidence than for the $V_{Coma}=0$ 
case given above.

\section{Discussion}

The magnitude of the bulk flow observed by Lauer and Postman
presents a serious challenge to models of structure formation.
However, if such a flow actually exists, its large amplitude
should make it readily evident in any peculiar velocity survey
that probes similar scales.  By studying a sample of galaxies
covering three regions in a plane roughly parallel to the LP
bulk flow, we have measured two components of the bulk flow on
similar scales to the LP survey.  Our sample galaxies show no
evidence for a large bulk flow.  Instead, our results indicate
a component of the bulk flow in the sample plane that is
consistent with expectations for a LDCDM power spectrum and that
lies $110^\circ$ away from the the LP flow.

A further analysis shows that the difference between the observed bulk
flows cannot be accounted for by the effects of measurement errors,
incomplete cancelation of smaller-scale velocity modes, or by the
choice of the zero point in the distance relation.  Indeed, if we
assume an LDCDM model for the power spectrum, we find that for our
results the LP bulk flow can be excluded at the $99.8\%$ level.
Since we consider the difference of the bulk flows of two surveys of
comparable scale, invoking extra power on scales equal to or larger
than the surveys will not alter this result.  The observed bulk flows
could be reconciled if we have vastly underestimated the power on
scales smaller than the surveys, but this would be in conflict with
other observations.

\acknowledgments       

GW wishes to acknowledge partial support from the Alexander von
Humboldt Foundation during a visit to the Ruhr-Universit\"at Bochum,
and also from NSF grant AST93-47714. KM acknowledges financial support 
from a Dartmouth Fellowship and an ESO Research Studentship, and is 
grateful to Roberto Saglia for the use of his programs for profile 
fitting and seeing correction.

\newpage

\figcaption[bins.ps]{Averaged peculiar velocities (in the CMB frame)
of early-type galaxies in region 1 (upper panel), region 2 (middle panel)
and region 3 (lower pannel), after the correction for Malmquist bias.
The bin size is 2000~km~s$^{-1}$.
The horizontal dashed line shows the weighted bulk-flow fit
to the data, and the dotted line indicates the prediction corresponding
to the bulk-flow result of Lauer \& Postman (1994).
\label{bins}}

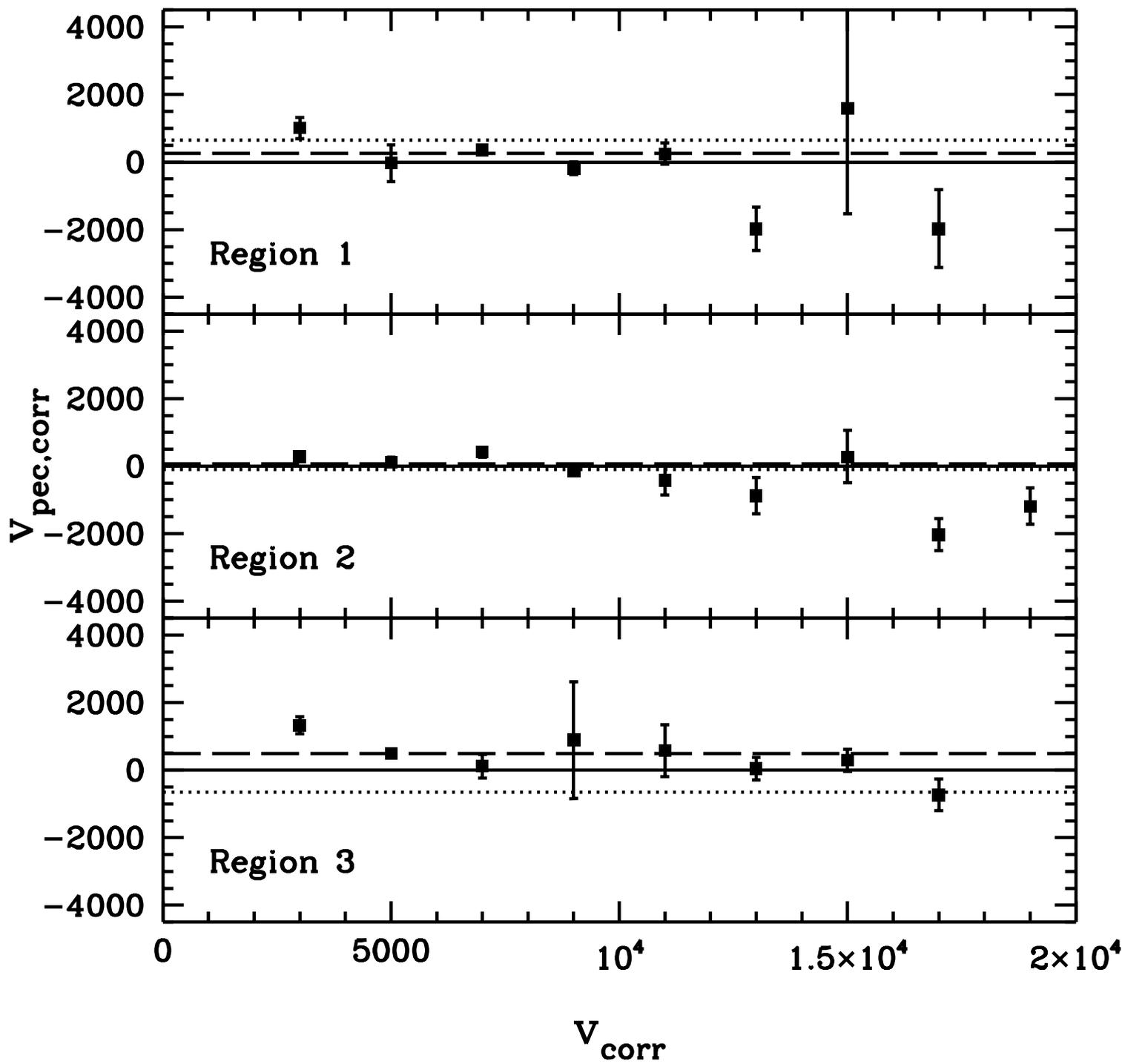
\end{document}